# Reflection-mode virtual histology using photoacoustic remote sensing microscopy

**Kevan Bell[1,2,†], Saad Abbasi[1,†], Deepak Dinakaran[2,3], Muba Taher[4], Gilbert Bigras[5], Frank K.H. van Landeghem[5], John R. Mackey[3], Parsin Haji Reza[1*]**

1. PhotoMedicine Labs, Department of Systems Design Engineering, University of Waterloo, Waterloo, Ontario, N2L 3G1, Canada
2. illumiSonics, Inc., Department of Systems Design Engineering, University of Waterloo, Waterloo, Ontario, N2L 3G1, Canada
3. Department of Oncology, University of Alberta, Edmonton, Alberta, T6G 1Z2, Canada
4. Division of Dermatology, Department of Medicine, University of Alberta, Edmonton, Alberta, T6G 2V1, Canada
5. Department of Laboratory Medicine and Pathology, University of Alberta, Edmonton, Alberta, T6G 2V1, Canada
†. Equal contributions.

*Corresponding Author:* phajireza@uwaterloo.ca

Contact information:

KB: k3bell@uwaterloo.ca

SA: srabbasi@uwaterloo.ca

DD: deepak.dinakaran@ualberta.ca

MT: mtaher@ualberta.ca

GB: bigras@ualberta.ca

FKHL: vanlande@ualberta.ca

JRM: jmackey@ualberta.ca

PHR: phajireza@uwaterloo.ca ; 1 (519) 888-4567 ext. 40172 ; E7-6416 200 University Avenue West, Waterloo, Ontario, N2L 3G1, Canada

## Abstract

Histological visualizations are critical to clinical disease management and are fundamental to biological understanding. However, current approaches that rely on bright-field microscopy require extensive tissue preparation prior to imaging. These processes are labor intensive and contribute to delays in clinical feedback that can extend to two to three weeks for standard paraffin-embedded tissue preparation and interpretation. Here, we present a label-free reflection-mode imaging modality that reveals cellular-scale morphology by detecting intrinsic endogenous contrast. We accomplish this with the novel photoacoustic remote sensing (PARS) detection system that permits non-contact optical absorption contrast to be extracted from thick and opaque biological targets with optical resolution. PARS was examined both as a rapid assessment tool that is capable of managing large samples (>1 cm$^2$) in under 10 minutes, and as a high contrast imaging modality capable of extracting specific biological contrast to simulate conventional histological stains such as hematoxylin and eosin (H&E). The capabilities of the proposed method are demonstrated in a variety of human tissue preparations including formalin-fixed paraffin-embedded tissue blocks and unstained slides sectioned from these blocks, including normal and neoplastic human brain, and breast epithelium involved with breast cancer. Similarly, PARS images of human skin prepared by frozen section clearly demonstrated basal cell carcinoma and normal human skin tissue. Finally, we imaged unprocessed murine kidney and achieved histologically relevant subcellular morphology in fresh tissue. This represents a vital step towards an effective real-time clinical microscope that overcomes the limitations of standard histopathologic tissue preparations and enables real-time pathology assessment.

## Introduction

Visualizing tissue pathology plays a central role in surgical oncology, cancer screening, drug development, and biological research. The standard histopathology workflow produces thin sections of tissue that are typically stained with dyes such as hematoxylin and eosin (H&E). H&E are the most widely used stains in anatomical pathology and are the long-standing gold standard for diagnosis and classification of cancer in tissue specimens from biopsies, surgeries and frozen sections. H&E are cationic/anionic based dyes that highlight specific sub-cellular contrast with hematoxylin staining the nucleus purple and eosin staining the surrounding cytoplasmic structures pink[1]. This process clearly accentuates the cellular morphology and enables diagnosis of pathology[2]. The preparation of histology slides, however, requires a potentially laborious multi-step process[3]. Tissue resected from biopsies or surgeries is typically fixed in formalin for up to 24 hours, then dissected. Representative samples are oriented, dehydrated (in which tissue water is replaced by alcohol, then xylene) and infiltrated with and embedded in paraffin wax to create a tissue block. The tissue blocks are sectioned with a microtome into approximately 4-5 micron sections then placed on glass slides. The paraffin is removed by a graded series of solvents, the tissue is then rehydrated, and finally stained with H&E. The slide is then interpreted by a pathologist using a transmission light microscope. Figure 1 illustrates the multiple steps of this process. This complex workflow can require at least 2 days to one week before a diagnostic report can be issued. The clinical turnaround time for complex specimens, such as radical cancer resections, may be greater than ten days for some cases[4]. Intraoperative tools such as frozen section analysis are commonly employed to guide surgical management and achieve negative resection margins, but this technique rapidly freezes the specimens and can introduce artifacts into the tissue morphology that could hinder clinical interpretation[5]. An ideal imaging method would produce H&E-like diagnostic quality images directly on unprocessed freshly resected tissue, which would save valuable time during biopsy assessment, permit more rapid intraoperative assessment of surgical margins, guide total resection of the tumor and reduce re-operation rates to excise involved surgical margins from cancer left behind.

A microscope that can disrupt the standard 100-year-old histopathology workflow and provide H&E-like contrast *in-situ* could radically change the clinical pathology paradigm and provide access to new data which could improve patient outcomes. However, a microscope suitable for clinical or research settings must meet several requirements: (I) The device must be capable of producing H&E-like diagnostic quality images. Pathologists are accustomed to assessing H&E stained tissue and cellular morphology. Thus, such a device must visualize cellular structures with good resolution and chromophore-specific contrasts, subsequently combining this contrast to generate images that resemble H&E stained slides which are readily interpretable. AI recognition systems that have previously been trained on conventional histology preparations can also be employed to detect cancerous regions in images from this device and potentially guide surgical decision-making or improve cancer screening. (II) The device must be capable of reflection-mode imaging. It is typically challenging for transmission-mode microscopes to visualize morphology on thick specimens such as freshly resected tissue or *in-situ* applications. This is primarily because the transmitted light is attenuated significantly as it travels through the thick opaque specimen. (III) The device must be capable of label-free visualization of intrinsic endogenous contrast. Exogenous dyes can be toxic and may require additional safety measures in clinical or surgical environments. (IV) The microscope must not require contact with the target in order to reduce the risk of infection and permit a rapid disinfection process between cases. (V) Real-time imaging would provide immediate feedback during surgeries and confirm suitability of tissue acquired in biopsy procedures. (VI) The device must be capable of 3-dimentional imaging or optical sectioning. Optical sectioning provides a means to visualize multiple layers of diseased tissue without the need for physical sectioning. (VII) Finally, it would be desirable if the microscope were able to image specimens at each intermediate step (as shown in Figure 1) during the standard histopathological process. This would enable parallel integration into existing workflows at hospitals and encourage adoption. These capabilities, when combined, would result in a microscope that is suitable for intraoperative environments and would facilitate diagnostic quality H&E-like contrasts in fresh tissue specimens or *in-situ*.

A variety of techniques have been developed to provide an alternative to standard histopathology. These methods have yet to widely adopted because they do not fully address the requirements described above. Techniques such as fluorescence microscopy[6-9] structured light microscopy[10,11], light-sheet microscopy (LSM)[12] and microscopy with ultraviolet surface excitation (MUSE)[13,14] have demonstrated promising results in providing H&E-like contrast on tissue mounted on microscope slides or freshly excised tissue. However, these methods cannot image unstained tissue and require the application of fluorescence dyes to the sample, adding time, expense, and the potential for occupational exposure to these chemicals. LSM has reported rapid volumetric imaging of unfixed tissue but requires additional processing steps such as optical clearing of the samples in addition to fluorescence dyes[12]. Optical coherence tomography (OCT) has been used for virtual H&E imaging with the method reporting cellular scale resolutions[15,16]. However, as OCT uses optical scattering to provide contrast it is not capable of easily differentiating among chromophores due to lack of specificity. Since H&E staining is chromophore specific, OCT images do not typically resemble H&E slides requiring pathologists to be retrained to interpret OCT visualizations[17,18]. Stimulated Raman scattering (SRS) modalities allow label-free optical imaging based on vibrational spectroscopy[19]. These devices have primarily been shown in a transmission-mode architecture, limiting the tissue to thin sections. SRS devices configured for reflection-mode operation employ specialized photodiodes between the objective and the sample[20]. Transmission-mode SRS microscopes have demonstrated H&E-like contrast in thin and unfixed tissue specimens without the use of exogenous dyes[21,22]. However, thick tissue was imaged by tightly squeezing the sample between coverslips which is unsuitable for imaging large specimens or directly imaging a resection bed[23]. In addition, the squeezing procedure may place considerable pressure on the sample, potentially damaging cellular morphology, interfering with accurate margin assessment due to distortion, and hindering the analysis of the sample with standard histopathological techniques.

Photoacoustic (PA) imaging techniques such as optical-resolution photoacoustic microscopy demonstrate exceptional visualizations of nuclear and cytoplasm morphology[24-26]. PA imaging takes advantage of the endogenous optical absorption contrast present within tissue. However, to detect this endogenous contrast PA devices employ an ultrasonic transducer in contact with the target. The requirement for contact with the acoustic transducer poses significantly higher risks and logistical challenges in maintaining surgical field sterility. The transducers are typically bulky devices which make their application in space-constrained resection sites difficult. Moreover, most PA methods for histology-like imaging have been demonstrated in transmission-mode which makes them unsuitable for unfixed tissue or *in-situ* imaging. PA imaging devices have previously used ring-shaped transducers for high resolution reflection-mode approaches. However, such devices require the tissue and parts of the apparatus to be submerged in water for effective acoustic coupling making them unfeasible for *in-situ* imaging[27]. These limitations, therefore, pose significant challenges for clinical or intraoperative applications.

Photoacoustic Remote Sensing (PARS™) is an emerging non-contact photoacoustic imaging technique[28,29]. PARS circumvents the limitations of conventional PA techniques by replacing the acoustic transducer with a continuous-wave detection laser[30,31]. This laser provides an all-optical design which allows for reflection-mode non-contact label-free imaging. These capabilities, as highlighted earlier, lead to a device suitable for intraoperative and clinical environments. Previously, PARS has demonstrated the ability to visualize nuclear contrast in chicken embryos and animal tissue[32,33]. However, these reports were limited in their field of view and did not reveal significant structural information. We recently demonstrated the visualization of nuclear morphology and hemoglobin contrast in human tissues in formalin-fixed paraffin embedded (FFPE) thin sections and blocks[34,35]. In the current study, using a multiwavelength PARS microscope, we present the first non-contact reflection-mode label-free device capable of visualizing nuclear morphology as well as the surrounding cytoplasm contrast, yielding images, similar to standard H&E prepared slides imaged with a transmission light microscope. We demonstrate the maturity of this technique by producing H&E-like diagnostic quality images of human brain tissue in FFPE unstained sections, human breast in FFPE blocks and human skin in unstained frozen sections. Finally, to demonstrate the ability to visualize H&E-like contrast in thick specimens, we image unprocessed freshly resected murine tissue. It is important to note that a single device imaged these sample types without modification, demonstrating technical versatility. Figure 1 illustrates the wide variety of sample types that PARS can image. Furthermore, improved image speeds makes whole slide imaging feasible and permits rapid gross assessment to find regions of interest to examine more closely. We acquire rapid single-wavelength large field of views and then provide high-resolution H&E-like images in clinically relevant regions in each of these sample types. These images are qualitatively compared to adjacent sections subject to the standard histopathological process. In aggregate, these results position PARS to challenge existing histopathological workflows, augment existing techniques or potentially superseding them in the future.

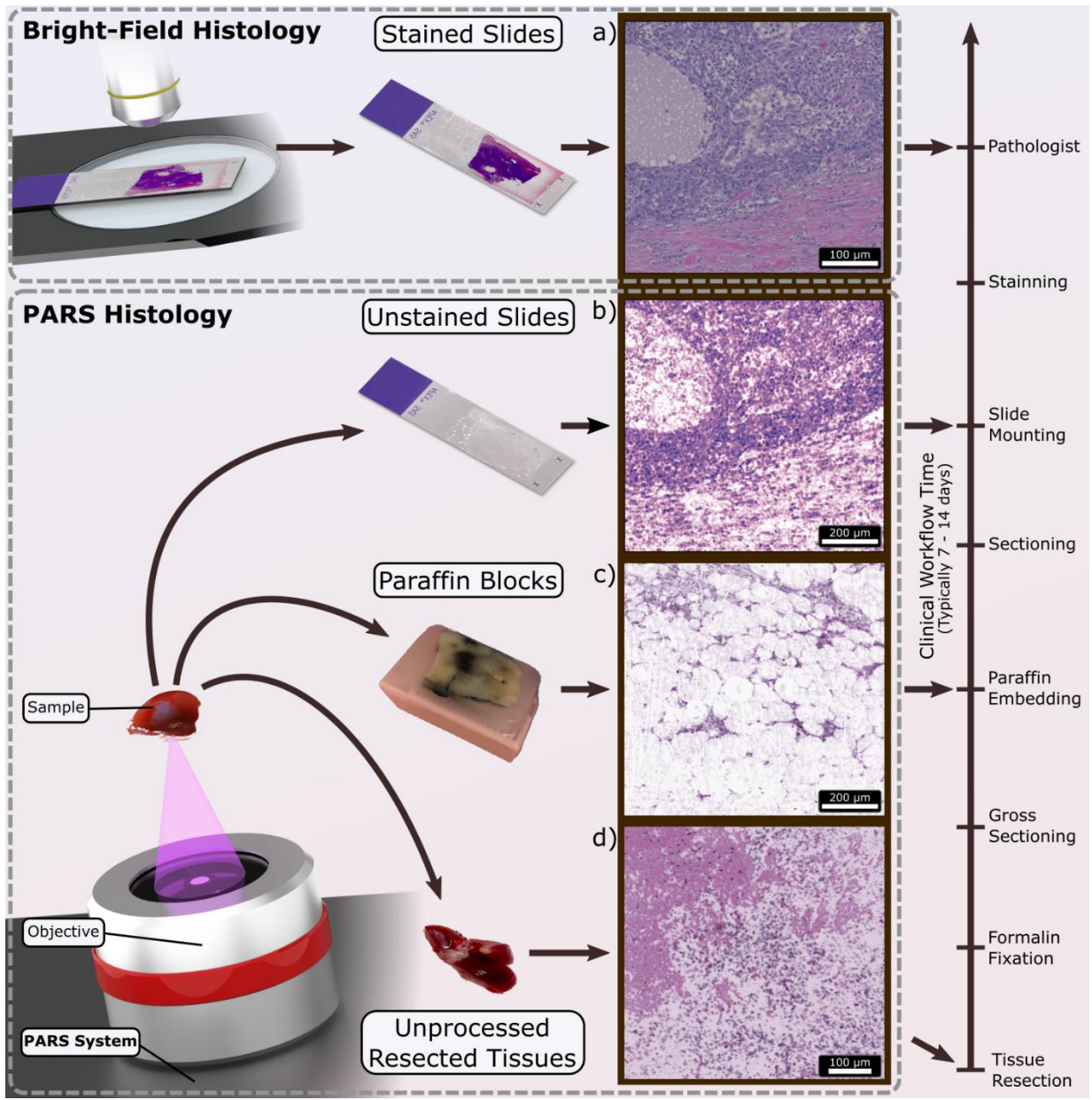

**Figure 1.** Overview of PARS histologic imaging workflow as compared to conventional light microscopy. (a) Conventional imaging of H&E-stained slides is performed on a bright-field microscope where the Hematoxylin (purple hues) and Eosin (red hues) stains block light from a white source. PARS may image (b) unstained FFPE slide preparations, (c) unstained FFPE blocks and (d) unprocessed tissues by taking advantage of the intrinsic optical absorption provided by the cell nuclei (DNA) and the surrounding cytoplasm (cytochrome). We image each intermediate step along the FFPE process in this paper using a single system configuration to show the versatility of PARS. No other reported technique has reported all of these capabilities in a single modality.

# Results

In this paper, for the first time, we show the unique features of the PARS technology in providing tissue imaging comparable to that provided by conventional H&E-stained FFPE preparations. H&E differentially stains tissue based on ionic polarity. Hematoxylin stains cationic regions like DNA in the nuclei as purple and eosin stains anionic regions such as cytoplasmic macromolecules in pink. To provide similar information, in this study, we report a PARS system that targets the ultraviolet absorption peak of DNA (~260 nm) to extract nuclear contrast and the blue light (420 nm) absorption of cytochrome to extract cytoplasm contrast. Details regarding wavelength selection are included in Supplement Information 1 and 2. This H&E-like contrast is shown in a two-color system which uses a tuneable nanosecond excitation source. However, this source only provided a relatively slow repetition rate (1 kHz) making it inappropriate for demonstrating rapid large sample grossing scans (>2 mm$^2$). To demonstrate the wide grossing capabilities of PARS, a faster repetition rate (50 kHz) single-color picosecond 266 nm laser was used. For brevity this article will refer to the single-color wide field-of-view (WFOV) system as imaging Mode 1, and the two-color multiwavelength system as imaging Mode 2. These architectures are highlighted in Figure 2 with further details provided in Supplementary Information 3. Both architectures use a continuous-wave 1310 nm detection source selected to maximize tissue penetration[36] which is co-aligned and co-focused with the excitation spot. Images are formed by mechanical scanning of the sample using a pair of scanning stages. Further details can be found in the Methods section. The characteristics of these imaging techniques such as resolution, sensitivity and imaging speed are reported in Supplementary Information 4, 5, and 6.

In order to help differentiate between the two systems, single-color acquisitions (Mode 1) taken with the 266 nm picosecond laser are presented in grayscale, and multi-color acquisitions (Mode 2) taken with the tunable laser are presented with a custom color map emulating H&E staining. The exact wavelengths to target in order to best simulate the contrast provided by conventional H&E is explored in the following sections. As well, to maintain a consistent system across varied sample types, the imaging head is set up as an inverted microscope so liquid and malleable solid samples can be easily flattened against a viewing window. This orientation also helps to level the tissue surface on FFPE blocks.

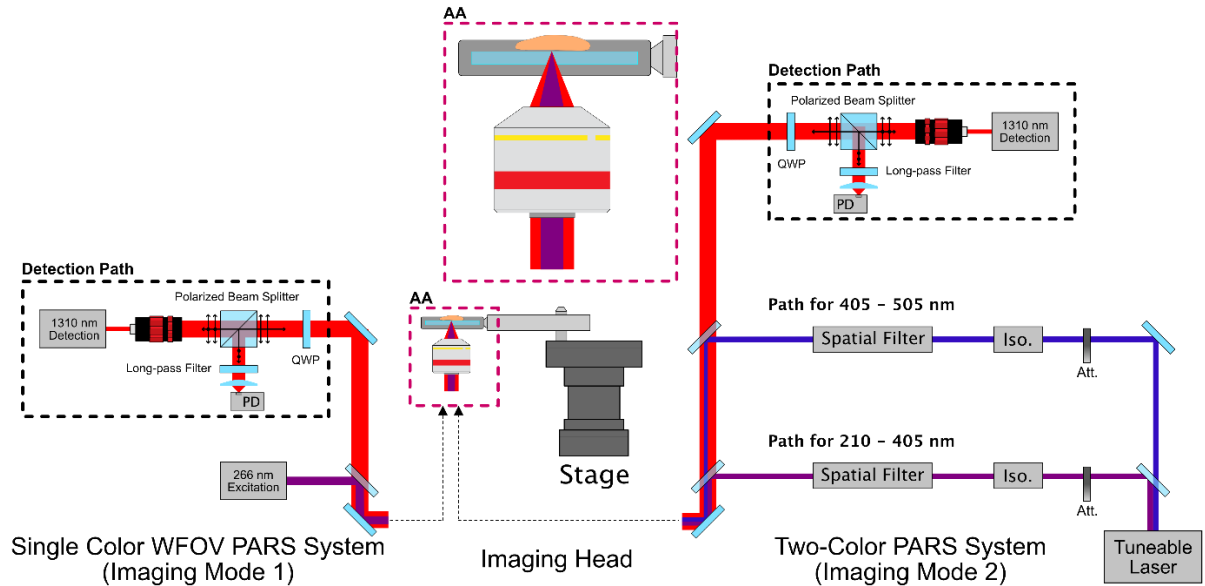

**Figure 2.** A simple system diagram showing the two excitation pathways. Component labels are defined as photodiode (PD), quarter wave-plate (QWP), Attentuator (Att), Isolator (Iso.).

PARS images generated for this work are produced in two different approaches depending on the system with which they were captured. WFOV single-color images were acquired using only a 266 nm excitation source and lack any inherent cytoplasm contrast. As such, a gray colormap is applied to the data with the aim of providing high contrast morphological information. Captures provided by the two-color system acquire both at 250 nm for nuclear contrast and 420 nm for cytoplasm contrast. A separate color map is applied to each of these two acquisitions to independently emulate the effects of H&E stains as they would appear on a conventional bright-field microscope. These two individual images are then combined through cyan magenta yellow and black (CMYK) color-space addition to produce a false-colour image. Colormaps are tuned such that they provide a close analog to matching H&E-stained bright-field images. Additional technical information on image formation is provided in the Methods section.

The first tissue preparation method explored in this work is that of FFPE slides. These paraffin-embedded samples are thin (roughly 4 µm) sections adhered to standard borosilicate glass slides and are prepared without a cover slip. A cover slip is not used as standard borosilicate glass cover slips are not transparent in 250 nm or 266 nm light. Slides are placed within the imaging head support with the exposed sample facing down towards the objective. Apart from the lack of any stains, this sample type is quite similar to the conventional H&E slides which would be viewed under bright field microscopes. They are flat and highly transparent without visible pigmentation. Here, we present human brain samples.

FFPE slide preparations of human brain tissues are presented in Fig. 3. Gross imaging of the entire specimen is provided by the single-color 266 nm WFOV PARS system (Fig. 3a) in a 21 mm x 13 mm acquisition. The adjacent slide underwent conventional H&E-staining for comparison, a large section of which is shown in Fig. 3b and its relative location on the PARS WFOV is shown in red. Two-color acquisitions were then performed on smaller regions (Fig.3c and 3e) at 1.6 mm x 1.6 mm with a 900 nm step and 300 µm x 300 µm with a 300 nm step respectively and are presented with their adjacent H&E counterparts (Fig. 3d and 3f). The false-

H&E colormap was tuned to match the bright field images. In the 1.6 mm field, PARS demonstrates its ability to recover the sparse nuclear structure within this healthy human brain sample. At this scale, further diagnostic qualities are accessible such as internuclear spacing and nuclear volumes. These features allow not only the differentiation of normal brain tissue, grey and white matter but also identification of specific cell types such as neurons, glial cells or endothelial cells. Fig. 3e highlights the current resolution of the two-color system with FFPE slides where nucleoli are visible within several nuclei. This resolution regime, and perhaps higher resolution still, is effective at visualizing intranuclear morphology. This aspect may be particularly beneficial when attempting to differentiate benign versus malignant pathology within a tissue sample. The 420 nm contrast provided additional morphology information such as the location of blood vessels through high concentrations of erythrocytes and is providing additional structural information within the internuclear regions. These results represent the first reports of multiwavelength photoacoustic acquisitions (non-contact or otherwise) which have produced dual contrast H&E-like visualizations of human tissues.

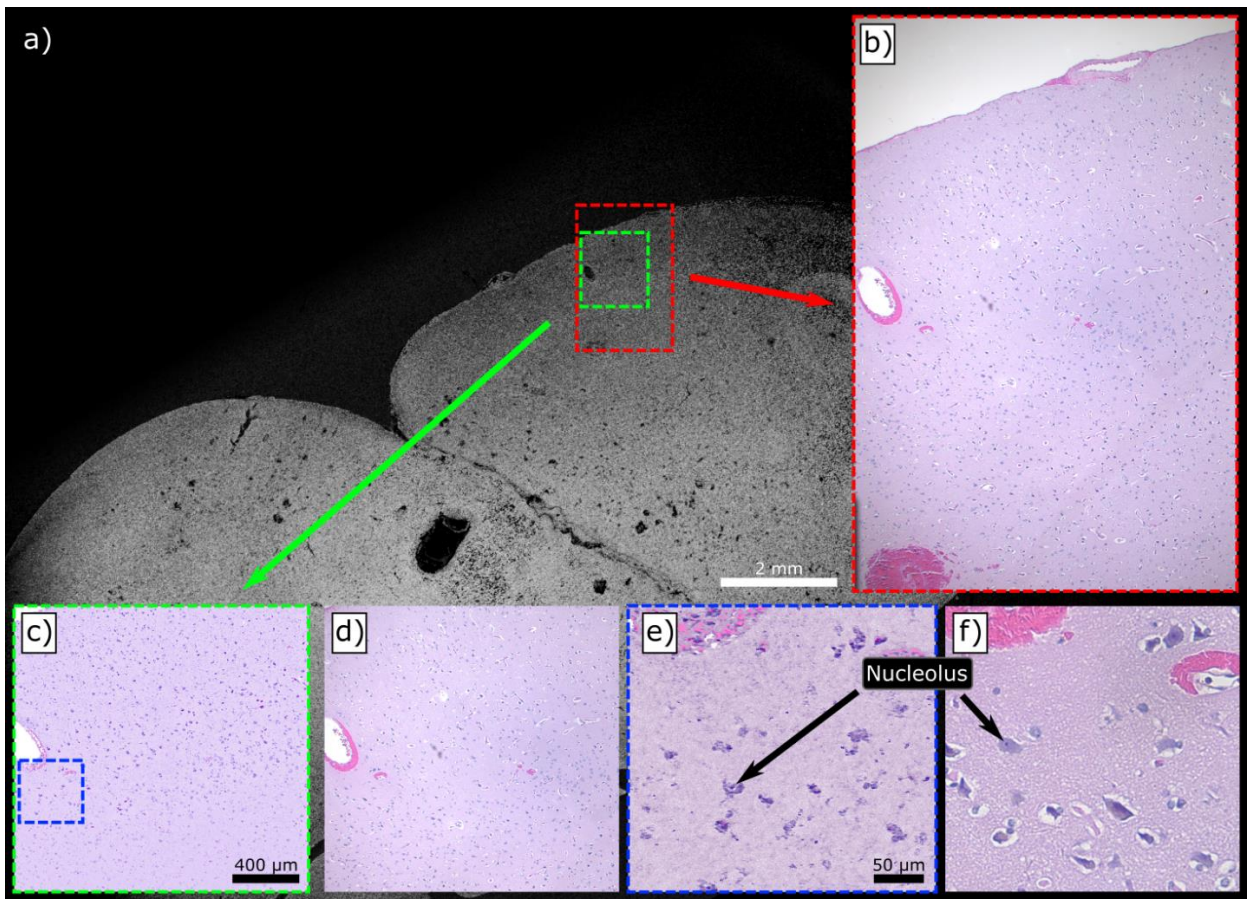

**Figure 3.** Several comparisons between PARS and conventional bright-field images of FFPE slides of human brain tissues. a) A WFOV scan using 266 nm excitation with b) a matching wide field image of the adjacent slide which has been H&E stained. c) A two-color (250 nm and 420 nm) PARS with a false-colour map applied to match d) the adjacent H&E region. Finally e) and f) likewise show a two-color PARS and bright-field image respectively in higher detail.

FFPE tissue blocks of human breast tissue with ductal carcinoma in situ (DCIS) were imaged next. The standard management of DCIS is breast conserving surgery which aims to remove the minimum amount of healthy tissue. About 20% of patients require secondary resections to convert resections with DCIS-involved surgical margins to DCIS free margins[37]. Tissue blocks from breast tissue resections are typically 3 mm thick and highly opaque, making transmission-mode modalities impractical. Blocks imaged here had adjacent sections removed by a microtome, leaving a flat surface for the PARS microscopes to image. These blocks were imaged inverted in with the tissue face down to ensure it was parallel to the plane of scanning.

Figure 4a shows a WFOV of the sectioned surface. This field of view encompasses almost the entire tissue contents of the block at 17 mm x 17 mm with a 4 µm spatial sampling. A smaller section of the block surface was then scanned at a 1 µm spatial sampling (Fig. 4b). At this level adipose tissue, stromal elements and regions of DCIS are clearly visible. A 3-dimensional volumetric scan of the region was also conducted with the single-color acquisition showing the optical sectioning capabilities and depth information which can be extracted (Fig. 4c). Two smaller scans were then performed at various locations on the block surface using the two-color PARS (Fig. 4d and 4e) with an 800 µm x 800 µm field-of-view (FOV) and 900 nm step. For comparison, inset into these figures are bright-field H&E images of similar regions. Here, the difference in contrast between the fat cells and the surrounding tissues becomes more apparent with the addition of the 420 nm hemeprotein contrast. The results also demonstrate that PARS can recover fine nuclear structure from within these FFPE blocks. This represents the first time that large sections of FFPE tissue blocks have been visualized by taking advantage of the intrinsic optical absorption contrast of DNA. When combined with the high imaging rate, these capabilities facilitate future development of a PARS-based rapid FFPE block grossing tool.

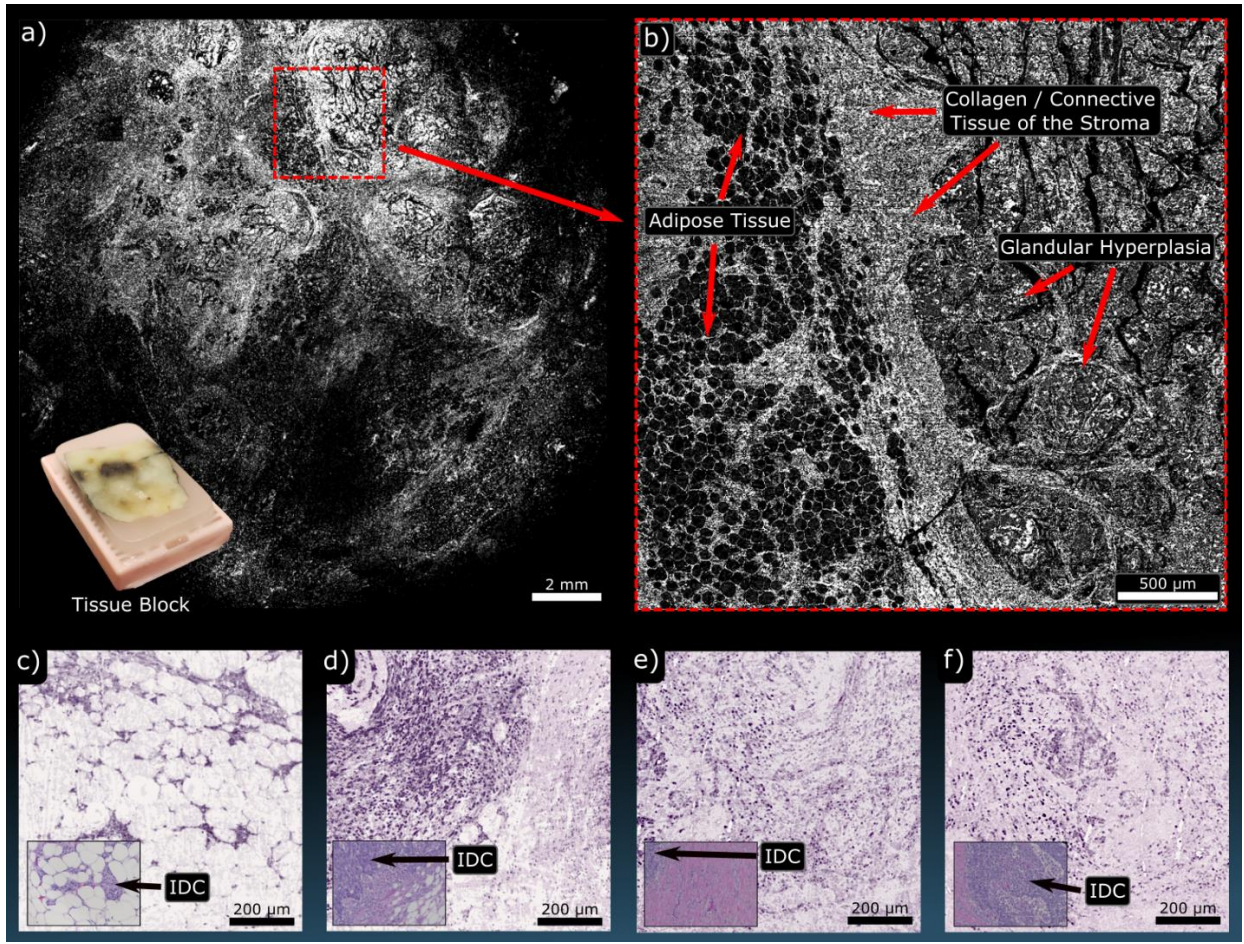

**Figure 4.** PARS imaging performed on FFPE tissue blocks of human breast. (a) highlights a WFOV single-color 266 nm acquisition covering nearly the entire tissue block surface. Inset is an image of the tissue block which was imaged. (b) shows a higher resolution single-color 266 nm of the highlighted region in red. At these scales bulk tissue components can be identified such as adipose tissue and fibroglandular tissue. (c-f) shows several two-color acquisitions from FFPE tissue blocks of human breast with brightfield images of similar H&E-stained regions inset. These all highlight regions of invasive ductal carcinoma (IDC).

Continuing with the same FFPE human breast tissue blocks, volumetric acquisitions were performed on the single-color WFOV system (Fig. 5). These were acquired by taking consecutive 2D scans at 5 mm x 5 mm with 4 µm lateral sampling. Depth scanning was then performed using a third mechanical scanning stage moving perpendicular to the surface. At various depths in the tissue block different bulk structures can be seen demonstrating the optical sectioning capabilities of the PARS technique. As well, since each depth level is separated by around 4 µm, each represents morphology which would be recovered on a single FFPE slide. As such, this technique may be useful in providing rapid virtual sectioning of the tissue block negating the need for multiple slides.

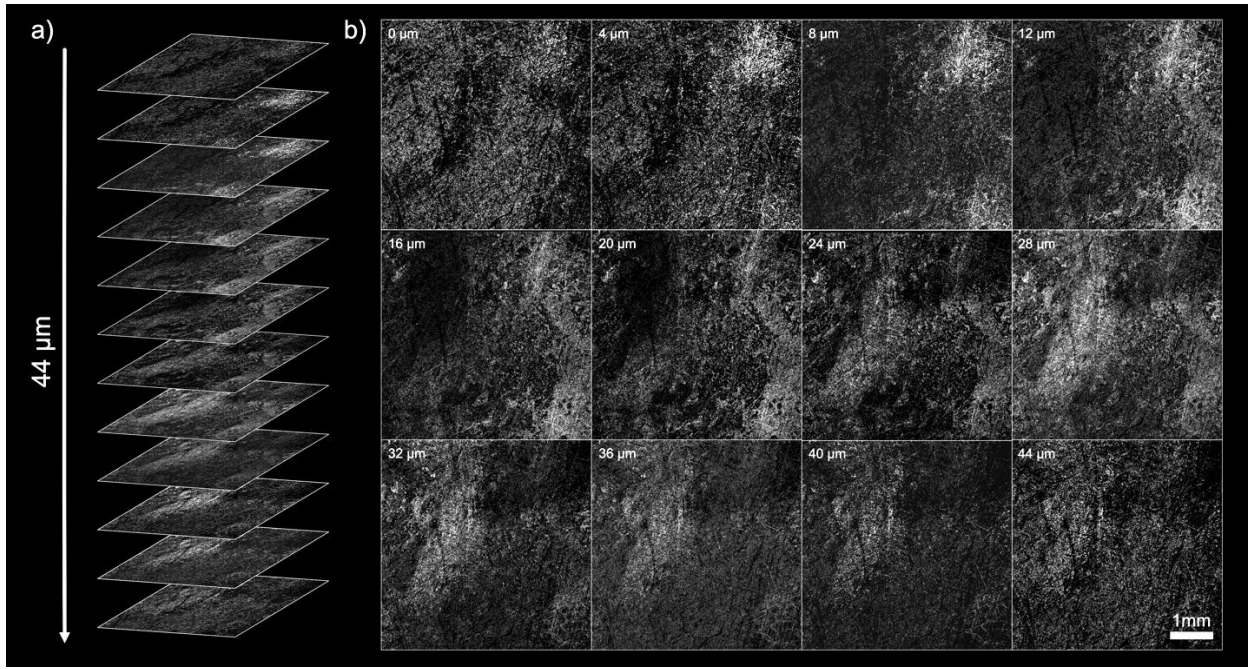

**Figure 5.** Several 2D sections from a 3D PARS scan of a FFPE human breast tissue block. (a) Shows the various slides which constitute the volume in a stack. (b) Shows several of the sections in greater detail.

Frozen sectioning is a tissue preparation system commonly performed intraoperatively to provide an initial diagnosis and assess the status of resection margins. Here, tissue is frozen, embedded within its cutting compound (optimal cutting temperature compound; commonly referred to as OCT) before sectioning on a cryo-microtome and staining for transmission light microscopy. This process allows for significantly faster turnaround time as compared to FFPE preparations, with frozen pathology visualizations being available within an hour in a clinical setting. However, this process reduces diagnostic accuracy and is therefore primarily reserved for oncological procedures where healthy tissue conservation is of importance[36]. For instance, this is the case for Mohs surgery, which is used to treat various skin cancers such as basal cell carcinoma (BCC) and squamous cell carcinoma (SCC)[38,39] in functionally or cosmetically important locations. Here, small sections of tissue are removed at a time and processed into frozen sections. The results from the frozen sections dictate the need for additional cycles of tissue removal and frozen section assessment. Since each cycle takes approximately one hour, a full Mohs resection can be a full day procedure and take significant time and technician resources. PARS may allow rapid and label-free visualization of these tissues at two separate stages. This would include imaging the frozen sections without the need for staining, which is explored in this article, or imaging the unprocessed tissue samples directly. As will be shown in the following section, PARS can recover cellular level morphology in unstained unprocessed murine tissue. Demonstration of its utility in unprocessed human tissue is planned.

Human normal skin samples are imaged with the WFOV system for gross analysis in roughly a 10 mm x 10 mm scan with a 4 μm spatial sampling (Fig. 6a). These scans take approximately 3 minutes to acquire at a 50 kHz interrogation rate. This rough scan helps to orient the sample within the imaging space. On this scale the various bulk layers of the tissue can be observed such as the epidermis, dermis and subcutaneous regions. Then, switching to the two-color PARS system, smaller FOVs are captured focusing on the epidermal layers with a 1.6 mm x 1.6 mm field and 900 nm spatial sampling (Fig. 6b and 6c). These are captured using 250 nm excitation for DNA contrast and 420 nm excitation for cytoplasm. Individual wavelength frames are combined to produce false color H&E-like images. From here, the tunable system was aimed to further enhance the epidermal layers moving to a smaller FOV of 300 µm x 300 µm field and 300 nm spatial sampling (Fig. 6c, 6d, 6e and 6f). From these higher resolution images, the sublayers of the epidermis begin to show including the outer stratum corneum followed by the granular layer, stratum spinosum and finally the basal layer below that (Fig. 6f). The dermo-epidermal junction is also clearly visible along with the dermal papillae (Fig. 6d).

Next, frozen sections were acquired from a Mohs micrographic surgery procedure removing BCC from a human subject (Fig. 7). Standard frozen pathology was done and sections were stained using toluidine blue rather than H&E as this represented a common processing for BCC. Additional adjacent image slides from the same resection block were taken to provide unstained examples for PARS imaging. Here, two WFOV single-color 266 nm PARS are presented which showed the entire frozen section with 4 µm lateral sampling (Fig. 7a and 7b). Figure 7c shows a higher density single-color scan of the highlighted region in Fig. 7b with 1 µm lateral sampling. This resolution level clearly resolves finer bulk structure and shows individual cell nuclei. Two-color acquisitions were then performed on the highlighted regions in Figure 7a (Fig. 7d and 7f). These can then be compared with the adjacent sections stained in toluidine blue (Fig. 7e). Here, the overall morphological similarities between the bright-field image of the stained slide and the PARS image of the unstained slide is evident. Much of the finer detail in the zoomed-in regions can be seen, with morphological features of normal skin tissue versus cancerous tissue being appreciable (Fig. 7d). These results highlight the rapid diagnostic potential of the PARS technique working with frozen sections.

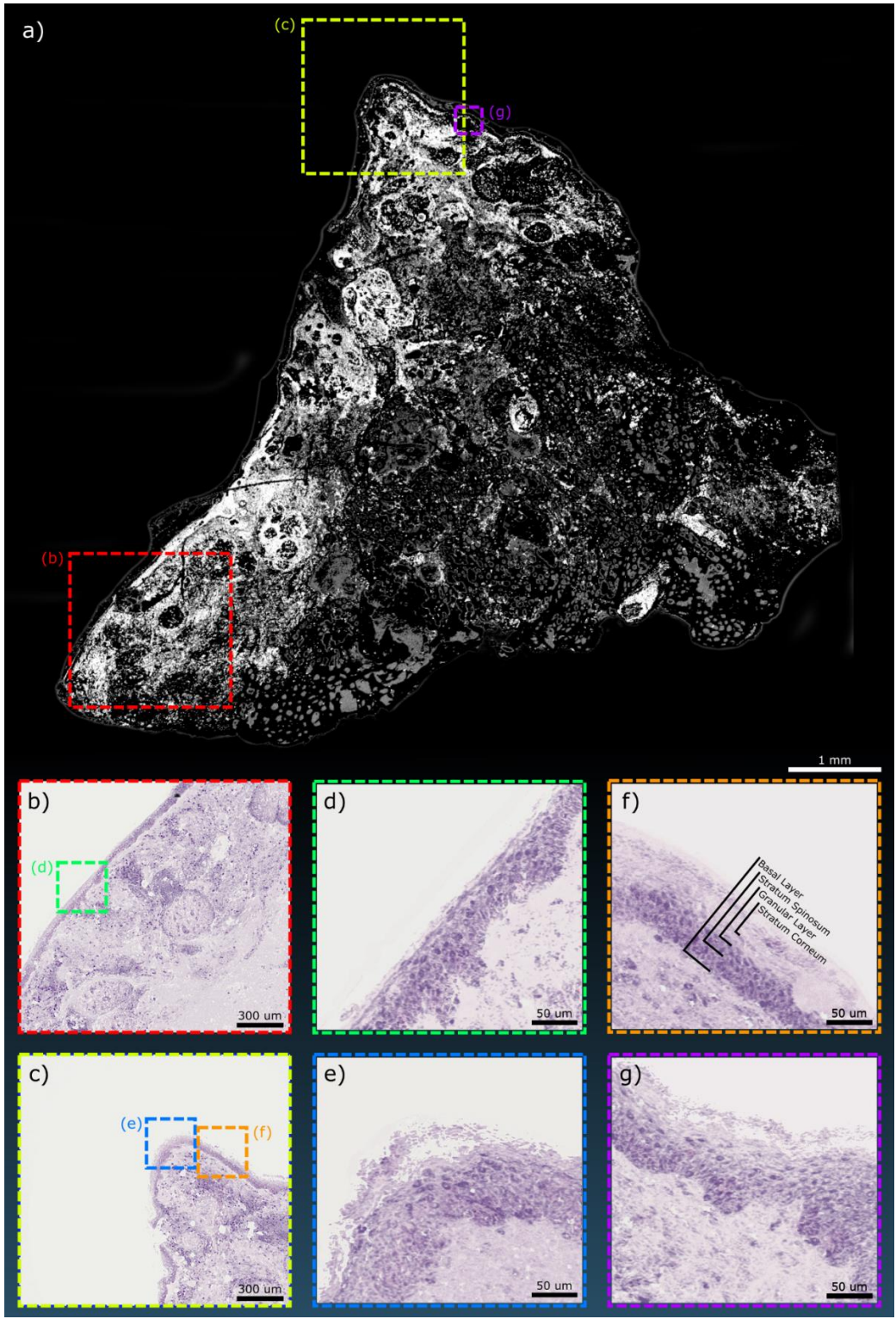

**Figure 6.** Several PARS images of a human skin sample mounted as a frozen section slide. (a) A WFOV PARS acquisition of the sample using the single color 266 nm system. The two-color PARS was then used over smaller field the views in (b) and (c) focusing on the outer tissue layers. Still smaller field of views are shown in (d-g) highlighting the details available within the epidermal layers. These layers are annotated in (f).

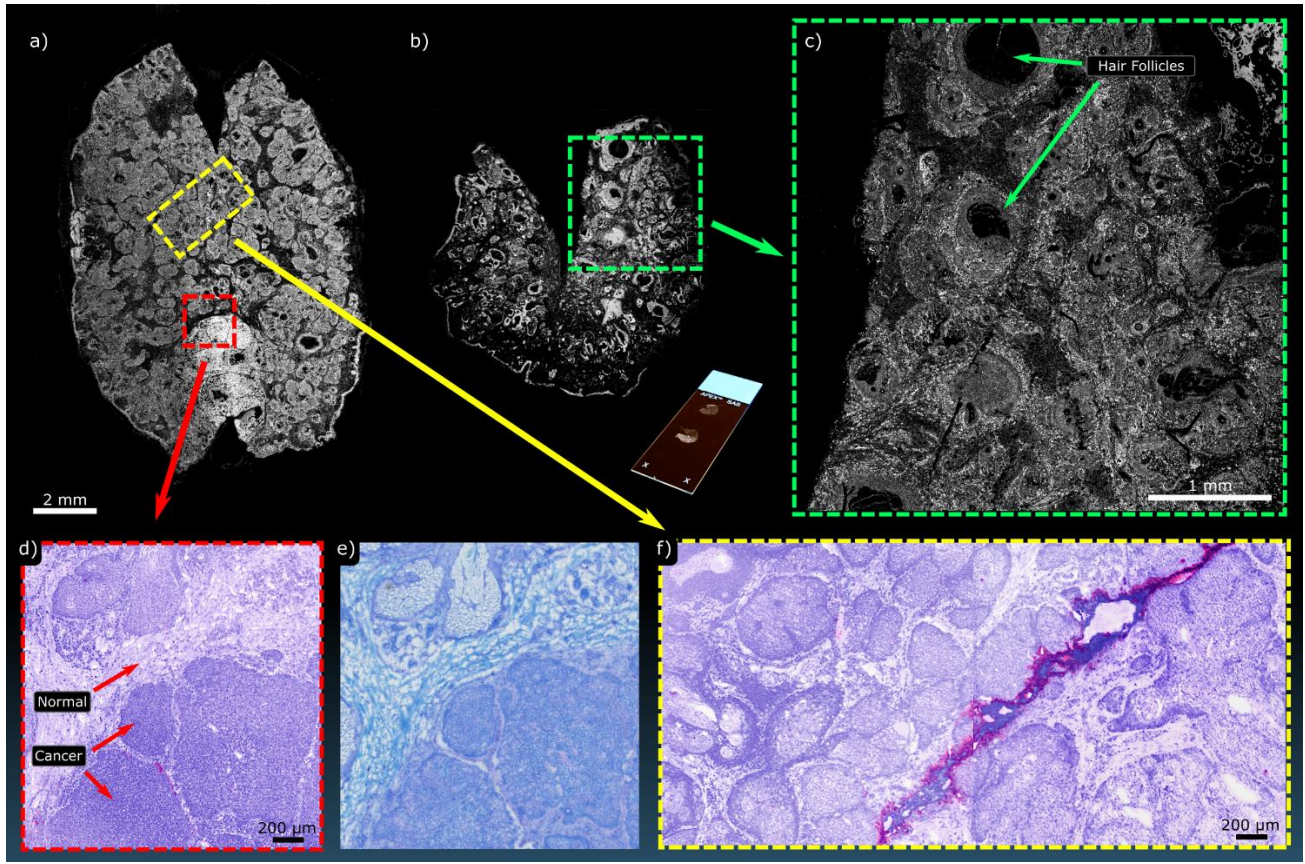

**Figure 7.** PARS imaging performed on frozen sections from a Mohs procedure. (a) and (b) show WFOV single-color acquisitions of two separate entire frozen sections. Inset with (b) is an image of the unstained section mounted on a glass slide. (c) shows a higher density scan of the highlighted region in (b). (d) shows a smaller FOV of the highlighted region in (a) captured with the two-color system along with (e) the corresponding section stained with toluidine blue captured on a standard bright-field microscope. (f) shows a region of subcutaneous healthy tissue captured on the two-color PARS which was likewise taken from the highlighted region in (a).

Fresh, unprocessed tissue was the final preparation studied in this work. Imaging of unprocessed resected tissues represents the penultimate goal for the PARS technique, superseded only by the ultimate goal of direct imaging of the surgical resection bed. By visualizing freshly resected tissue, PARS could provide rapid clinical feedback eliminating the need for further tissue processing. Imaging bulk tissue samples presents new challenges, including the uneven surface which may be left from the removal process. This requires either tracking of the sample surface (to be addressed in future works) or placing the sample against a window to provide a flat viewing surface suitable for standard PARS microscopy. Another challenge arising from unprocessed samples is the limited time available to image prior to tissue degradation after devitalization. For the purpose of this initial investigation, samples were imaged within 3 hours of resection to minimize tissue degradation.

Unprocessed murine kidneys were transported in room temperature PBS for transport and cut to produce sagittal sections which were then placed with the fresh cut against the UV viewing window of the system. A WFOV single-color acquisition captured nearly the entire organ (Fig. 8a). Several bulk features can be identified including the Calyces, Medulla and Cortex. These regions are highlighted within the figure. As well, several smaller regions were captured using the two-color system as shown in Figure 8b and 8c. The smaller regions were taken around the medulla. Imaging required careful removal of excess fluids and blood by washing with fresh PBS, as the transport PBS produced measurable PARS signal without discernible morphology. This signal is assumed to be protein and other macromolecules which could provide non-zero signal under UV and blue light excitation. Despite these challenges, for the first time, PARS has demonstrated H&E-like visualizations of unprocessed tissue morphology. These experiments were performed with a large gap between the sample and the objective (>7 mm) and without the use of any exogenous contrast agents. This represents a vital step towards PARS becoming an effective clinical tool as a method of rapid tissue assessment.

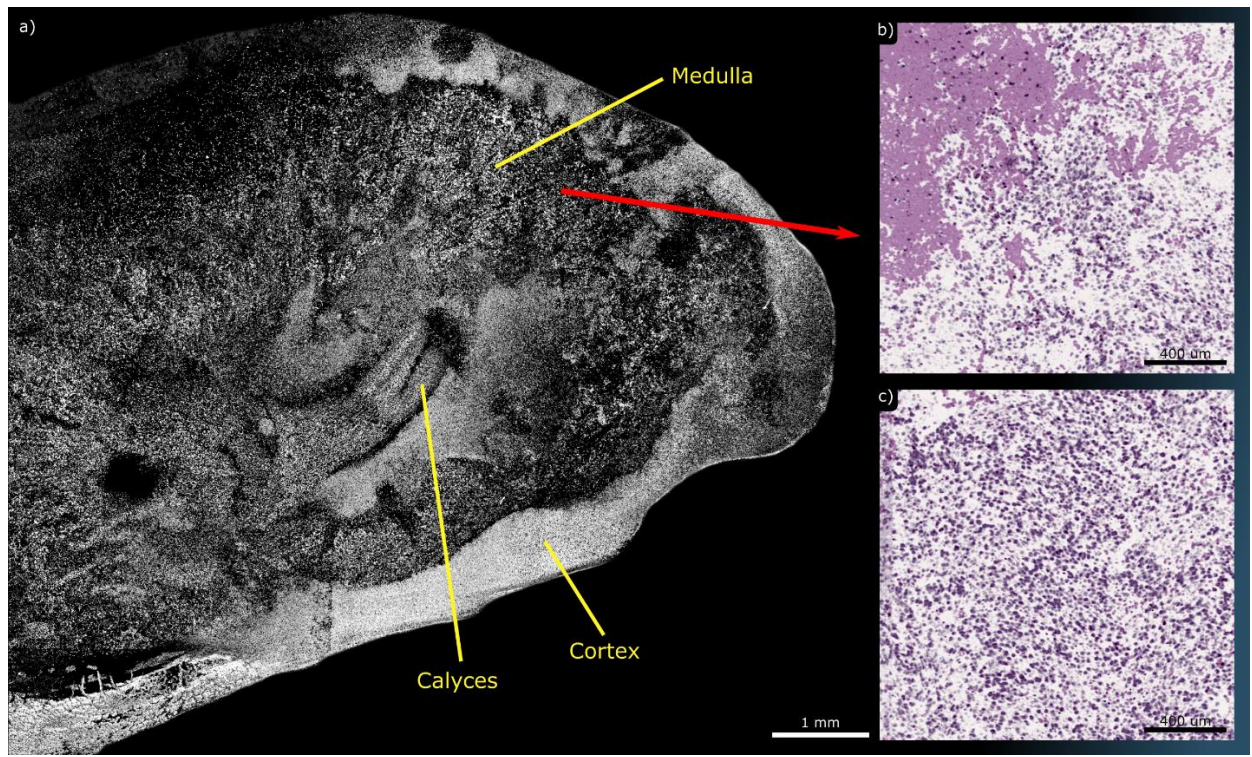

**Figure 8.** PARS imaging performed on unprocessed murine kidney. (a) shows a WFOV single-color acquisition of the organ which has been sectioned in half along the sagittal direction. Several bulk components are labeled. (b) and (c) shows smaller FOVs of the medulla region captured on the two-color system illustrating numerous tubule constituents of nephrons.

## Discussion

This work presents the efficacy of PARS microscopy in visualizing a wide variety of tissue preparations including human FFPE slides, FFPE tissue blocks, frozen pathology sections and unprocessed murine kidney. PARS was able to recover micron-scale tissue morphology solely by taking advantage of the intrinsic optical absorption properties of the tissue. Moreover, for the first time, the PARS technique has demonstrated true H&E-like contrast of both nucleic morphology and the surrounding cytoplasm. This achievement of non-contact label-free tissue contrast has thus far only been reported for coherent Raman microscopes which rely on costly ultrashort femtosecond sources and generally operate in transmission mode[23], limiting their utility to thin sections of tissue. The ability to image unstained fresh tissue directly may represent a practice changing technology for surgical management of cancer, which currently relies on lengthy processing methods. PARS microscopy functions both as a potential replacement for conventional bright-field processing techniques and as a rapid grossing tool capable of characterizing large regions of tissue. By presenting clinicians with rapid feedback within minutes, verified negative surgical margins can be achieved intra operatively much more quickly and efficiently, and reducing the need for additional surgeries. Furthermore, PARS maybe useful as a tool for virtual biopsy, to interrogate tissue prior to biopsy, as well as a rapid quality assurance step to determine if diagnostic quality tissue is obtained. Endoscopic configurations with PARS capability could reduce unnecessary biopsies and limit the load placed on pathology services, thereby improving patient outcomes and lowering healthcare costs. Finally, PARS interrogation of fresh tissue has the theoretical potential to identify biological features, such as lipid subtyping and cellular hydration, that are routinely lost in standard tissue preparation.

In this paper two separate excitation pathways were examined each with their own merits. The single-color WFOV PARS system performed rapid grossing of centimeter-scale tissue samples by using a 266 nm picosecond excitation source matched with a fast-mechanical scanning stage capable of a 300 mm/s scan velocity. This hardware facilitates scan rates of around 1.6 s/mm$^2$ at 4 µm steps allowing for large samples (>1 cm$^2$) to be visualized in under 8 minutes. Improved scanning methods, such as hybrid mechanical-optical scanning, adding galvanometer mirrors or polygon scanners, may allow a much faster interrogation rate with an order of magnitude improvement in imaging speed. If such a rate increase were accomplished entire tissue-block sized samples (~4 cm$^2$) could be image within minutes with sampling at ~4 µm. The two-color tunable excitation source operated at a far slower 1 kHz pulse repetition rate but allowed for exploration of optimal histological wavelengths. As well, this source facilitated the first non-contact reflection mode H&E-like images. Future systems for clinical deployment will use faster repetition rate non-tuneable sources at these optimal wavelengths. These sources may also facilitate multi-color WFOV H&E-like acquisitions at rates suitable for clinical use. Meanwhile, the tunable system remains a valuable tool for investigating additional contrast (lipids, melanin, histones, etc.). The current tunable system was relegated to smaller FOVs to maintain pragmatic imaging times. Per wavelength area scan rates were substantially slower as compared to the single-color counterpart with performances of around 21 min/mm$^2$ at 900 nm steps and around 178 min/mm$^2$ at 300 nm steps. The two detection pathways proved complimentary, with the single-color version demonstrating the potential imaging speed of the method when used for gross assessment, and the two-color version demonstrating available contrast and H&E-like visualizations.

We emphasize that no sample-specific changes to the PARS microscopes between sample types, and no special preparation was needed to any of the samples prior to imaging. This is yet another pragmatic consideration that makes PARS unique among competing modalities such as MUSE[13] or LSM[12], in that samples can be directly imaged in their standard form. However, the system did perform differently on different tissue preparation types. The FFPE slides and tissue blocks provided lower signal intensities and higher background signal levels, probably due to macromolecule degradation brought on by the process of formalin-fixation and embedding in paraffin, including denaturation and crosslinking of DNA[40]. The more ubiquitous background signal seen in the samples is likely to be a result of the paraffin within and surrounding the tissues. Paraffin provides a non-zero signal at all three wavelengths used. However, paraffin's lack of organized structure makes it easy to discern as background and separate from the tissue. In comparison, frozen section samples provided notably lower background and higher signal levels.

These would be attributed to reduced DNA damage providing additional contrast, and lower signal provided by the background embedding material. However, the highest contrast was observed in unprocessed tissues which completely lacked background signal and provided similar contrast to that seen in the frozen sections.

The resolution of both systems was fundamentally limited by their respective mechanical scanning stages. This limited achievable resolution to approximately one micron, obscuring some subcellular detail. Where higher resolution is desired, optical scanning or higher-resolution scanning stages may be required, with the resulting trade-off between spatial sampling density and imaging speed. However, these parameters can be optimized for each specific use. For example, in a Mohs procedure preference may be given to imaging speed as cancerous regions may be easily identified at lower resolutions of several microns. For applications requiring finer sub cellular details, such as for central nervous system pathology, a low-density grossing scan to orient the sample within the imaging space and to find regions of interest can be followed by smaller high-resolution scans. Since it is desirable to maintain Nyquist spatial sampling, such that sampling locations are separated by less than half of the lateral optical resolution, these choices also drive the selection of optics. For example, a high NA (such as 0.5) objective lens is inappropriate if the desired lateral spatial sampling is only 4 µm. Here a lower NA around 0.02 may be more appropriate to better match the spatial sampling. The ultimate power of the objective then dictates the necessary excitation required to maintain desired sample fluence.

The same trade-off could be said with respect to wavelength selection. If it is the desire to replicate multiple stains, such as H&E, multiple wavelengths must be used. Again, for some use cases single wavelength acquisitions may be enough, such as targeting DNA contrast in a Mohs procedure which is performed using a single cationic stain, toluidine blue. Other applications may require the additional contrast provided by two (H&E) or more excitation wavelengths. For example, if lipid contrast is desired in an unprocessed sample, SWIR such as 1200 nm may be added[41]. As more application-specific PARS investigations proceed in the future, these parameters must be appropriately optimized and balanced.

Our studies demonstrate the range of applications to which PARS, a novel tissue imaging technology, can be applied. By virtue of the multiple pragmatic advantages inherent to this technology, including non-contact, label free, cellular level resolution with multi-wavelength contrast, and rapid reflection mode image acquisition, PARS represents a vital step towards an effective real-time clinical microscope that overcomes the limitations of standard histopathologic tissue preparations and enables real-time pathology assessment. With configurations optimized to individual clinical applications, the PARS platform technology has the potential to improve diagnostic and therapeutic workflows in a variety of clinical settings.

## Methods (3000 words)

### Single-color WFOV PARS System

The single-color WFOV PARS system employs a 266 nm excitation laser with a 50 kHz repetition rate (WEDGE XF 266). This laser also outputs a 532 nm beam as a result of frequency doubling the primary 1064 nm wavelength. The 532 and 266 nm beams are separated using a CaF2 prism (PS862, Thorlabs Inc.) and the 532 nm beam is directed towards a beam dump. The 266 nm beam is then expanded using a variable beam expander (BE05-266, ThorLabs) and combined with the detection beam using a dichroic mirror (HBSY234, ThorLabs). The system utilizes a 1310 nm continuous-wave superluminescent diode detection laser (S5FC1018P, ThorLabs Inc.). This beam is polarized vertically and passed onto a polarizing beam splitter (CCM1-PBS254, ThorLabs Inc.). The polarizing beam splitter transmits most of the forward light towards the quarter wave plate. The quarter waveplate converts the vertically polarized light into circularly polarized light.

The two beams are then co-focused on the sample using a 0.3 NA reflective objective (LMM-15X-UVV, Thorlabs Inc.). Since the system is in an inverted configuration, the sample is placed on a UV-transparent optical window. The optical window is resting in a circular holder which is connected to the mechanical stages (XMS100-S, Newport Inc.) as shown in Figure 2. The back-reflected light from the sample is then converted back to linearly polarized light using the quarter wave plate and directed towards the polarizing beam splitter. The polarizing beam splitter then reflects the majority of the back-reflected light towards the photodiode (PDB425C, Thorlabs Inc.). A long-pass filter with a 1000 nm (FELH1000, Thorlabs Inc.) cut-off blocks any 266 nm back-reflection letting only the 1310 nm beam reach the photodiode. This light is then focused onto the photodiode using an aspheric condenser lens (ACL25416U, Thorlabs Inc.).

**Two-color Tunable PARS System**

The two-color PARS system utilizes a tunable source with a tuning range of 210 – 2600 nm (NT242, Ekspla Inc.). The beam from this excitation laser is split into different optical pathways depending on the wavelength. This particular study primarily uses 250 nm and 420 nm wavelengths. The laser light is first split using a dichroic mirror (HBSY134, Thorlabs Inc.) with wavelengths less than 405 nm being reflected and greater than 405 nm being transmitted. To prevent the laser from back reflection damage both wavelength ranges are passed through a beam isolator composed of a polarizer and quarter-wave plate. Wavelengths less than 405 nm are then focused into a pinhole using an achromatic doublet lens (ACA254-100-UV, Thorlabs Inc.) for spatial filtering. The filtered light is then collimated using a second doublet lens (ACA254-100-UV, Thorlabs Inc.) and passed through a beam expander (BE02-UVB, Thorlabs Inc.). The expanded light is combined with the detection beam using a dichroic mirror (HBSY234, Thorlabs Inc.).

Wavelengths greater than 405 nm are split again using a dichroic mirror with a 505 nm cut-off (DMSP505, Thorlabs Inc.). Wavelengths less between 405 - 505 nm are then spatially filtered, collimated and expanded similar to the less than 405 nm path. Wavelengths greater than 505 nm are spatially filtered by focusing the light into a single mode fiber. The light from the optical fiber is collimated and combined with the 405 – 505 nm path using a dichroic mirror (DMSP505, Thorlabs Inc.). The resulting optical path is combined with the detection beam using a subsequent dichroic mirror (DMLP1000, Thorlabs Inc.).

The tuneable PARS system uses a similar optical path for detection as the single-color WFOV PARS system. It employs a 1310 nm superluminescent diode (SLD1018P, Thorlabs Inc.) which is first passed through a polarizing beam splitter. The polarizing beam splitter transmits the forward light towards the quarter waveplate which converts the linearly polarized light to circularly polarized light. This light is then combined with rest of the optical paths using dichroic mirrors. The combined light is then focused onto the sample using a 0.5 NA reflective objective (LMM-40X-UVV, Thorlabs Inc.). The back-reflected light from the sample is then converted to the linearly polarized light using the quarter wave plate (WPQ10ME-1310, Thorlabs Inc.) and is reflected towards the photodiode (PDB425C, Thorlabs Inc.). A long pass filter (FELH1000, Thorlabs Inc.) ensures no excitation wavelengths influence the photodiode. The remaining 1310 nm light is then focused onto the photodiode using an aspheric condenser lens (ACL25416U, Thorlabs Inc.).

**Data acquisition**

The single-color and two-color PARS system use the same data acquisition scheme. The photodiode output is connected to a data acquisition card (DAC) (CSE161G4, Gage Applied). The sampling rate of the DAC is set to 200 MS/s to accommodate the 75 MHz bandwidth of the photodiode. The mechanical stages move the sample in a raster scan pattern specified in a position-velocity-time (PVT) file. These PVT files specify the trajectory for both the stages in the form of position, output velocity and time. The stage controller automatically calculates the acceleration and intermediate positions as a function of time to satisfy these three parameters. These PVT files also configure the stages to output a low signal when the fast axis reaches a constant velocity. When the stages are turning (i.e. are not at a constant velocity), this signal is high. This technique is used as the mechanical stages do not provide an method to probe or read the encoder positions for large. The constant velocity of the stages combined with the pulse repetition rate of the laser leads to point acquisitions at a fixed step size from each other. Therefore, the image reconstruction requires us to record two signals with the DAC, the PARS signal and the fast axis stage velocity signal.

Since PARS relies on initial pressure measurements the DAC is configured to record only a few hundred segments after each excitation event, typically only 128 segments. With a sampling rate of 200 MS/s, this results in a recording time of 640 ns. Large field of views often necessitate the acquisition of greater than 50 million points. The length of the point acquisitions combined with the number of require points result in a large amount of data that must be managed. For example, for a field of view with 50 million points, the memory and storage requirements for the raw 16-bit time-domains can be as large as 25 gigabytes, quickly making whole slide imaging impractical. To save on memory and storage requirements, only the amplitude of the PARS signals is stored alongside the fast axis' velocity signal. This results in significant memory savings and a 50 million point acquisition requires only 200 MB of storage. To acquire multi-wavelength images using the two-color PARS system the data acquisition software can be configured to acquire a series of images at various wavelengths. For example, for this study the system is configured to first image at 250 nm to acquire DNA contrast and then switched to 420 nm to acquire cytoplasm contrast as two consecutive acquisitions.

**Image reconstruction**

Since the data acquisition scheme is common to both the single-color and two-color PARS system, the image reconstruction is also very similar. The image reconstruction is primarily performed in a MATLAB script. This script relies on the trajectory file that was generated prior to a scan and the image data saved by the acquisition software. The image reconstruction script attempts to find locations in the image data where the fast axis' signal toggles from high to low and subsequently low to high. The indices between these transitions mark where the fast axis achieved constant velocity for each line. As these indices are located, the script copies corresponding indices from the PARS signals and stores them in a new array. The constant velocity of the stages combined with the excitation laser's constant repetition rate leads to these PARS signals being acquired a fixed step size apart. The trajectory file provides the stage's coordinates where each of these transitions from which the length of the line can be computed.

By dividing the length of the line by the number of trigger events per line, the step size between pixels can be computed. This enables the reconstruction script to simply plot the PARS signals in a grid leading to an image reconstruction without any interpolation.

**False-color formation**

Tissue stained with histological dyes exhibits rich colours and contrast. Stains colour different tissue components in patterns that a pathologist is trained to recognize and examine. To provide similar information to a pathologist, it is therefore necessary to colourmap the PARS images in a similar manner. H&E is the most commonly used staining media in histology with the hematoxylin staining DNA as purple and eosin staining cytoplasm as pink. To emulate these colours, an inhouse developed software maps the pixel intensities of the 250 nm image to a purple colour to simulate staining with hematoxylin. Similarly, the 420 nm image is mapped to a pink hue to resemble staining with eosin. Once the individual images are colorized, their saturation levels are adjusted and a gaussian smoothing filter is applied to reduce colour noise. The colourized images are then converted to CYMK color space as it was found to be more consistent across different images. This is likely a result of the CMYK-space subtractive nature being a close analogue to the transmission loss of light through dye mixtures. The CMYK images are then added together and converted back to RGB color space for display purposes.

**Sample acquisition, preparation and ethics review**

This study examines human tissue in three different sample types i) unstained skin and brain tissue sections on glass slides ii) breast tissue fixed in formalin and embedded in paraffin iii) frozen sections of skin from Mohs surgery. The clinical collaborators at the Cross-Cancer Institute (Edmonton, Alberta, Canada) obtained samples from anonymous patient donors and removed all patient identification from the samples. The samples were obtained under a protocol approved by Research Ethics Board of Alberta (Protocol ID: HREBA.CC-18-0277) and University of Waterloo Health Research Ethics Committee (Humans: #40275 Photoacoustic Remote Sensing (PARS) Microscopy of Surgical Resection, Needle Biopsy, and Pathology Specimens). All experiments were performed in accordance the relevant guidelines and regulations. In addition, freshly excised tissue from mice was obtained to demonstrate PARS' performance in imaging unprocessed tissue (Photoacoustic Remote Sensing (PARS) Microscopy of Resected Rodent Tissues; Protocol ID: 41543). The preparation method for all samples types is described below.

**FFPE Sample preparation**

To prepare FFPE blocks the tissue was submersed in formaldehyde for 48 hours. The tissues were then dehydrated by repeatedly immersing the tissue in ethanol of increasing levels of concentration ending in a 100% concentration of ethanol. The tissues were thereafter cleared with xylene to remove ethanol and any residual fat tissue. This clearing permits molten parrafin wax (60°+C) to penetrate the tissue. This embeds the tissue in paraffin wax. As the paraffin cools to room temperature, the FFPE tissue blocks are mounted in a cassette and completed. Brain and breast tissue specimens are prepared using this process. To further prepare unstained thin tissue slices on glass slides, 5 µm ribbons are sectioned using a microtome and placed onto glass slides. The glass slides are then baked at 60 °C for 60 minutes to remove excess paraffin from the sections. A comparison stained section for brain tissue specimens was obtained by immediately cutting the next ribbon (within 10 um), transferred to glass slides and then baked for 60°C for 30 minutes. The specimens were then stained with H&E contrast dyes and covered with mounting media and a coverslip. Once the mounting media was dry, the slide were fully prepared.

**Frozen Section Sample preparation**

The frozen sections for skin specimens with BCC were obtained via Mohs surgery. Tissue specimens of sizes up to 15mm x 30 mm are embedded in an optimal cutting temperature compound and placed in a cryostat pre-cooled to -20°C to -25°C. The specimens are then frozen to the pre-cooled temperature for 1 – 10 minutes depending on tissue components and thickness (ex: dermal tissue required closer to 10 minutes, fatty tissue requires about 1 minute). The frozen samples are then sectioned at a thickness of 5-10 microns and transferred to a warm (room temperature) microscopic slide. The slide is then air-dried and heat-fixed at 55°C for 1 minute. The sections are then stained with H&E to characterize the structure of normal skin, and with 1% toluidine blue aqueous solution which is a common staining protocol for BCC. Once the staining has been performed, the slides are cover-slipped with mounting media.

**Unprocessed Resected Tissue Sample preparation**

To demonstrate imaging on unprocessed tissue, unused murine kidney specimens were obtained with the aid of collaborators at the Central Animal Facility, University of Waterloo performing work under animal care approval (Photoacoustic Remote Sensing (PARS) Microscopy of Resected Rodent Tissues; Protocol ID: 41543). Kidneys were then excised and placed in PBS immediately. One of the kidneys were then cut longitudinally and imaged immediately using the PARS system after a 3-hour devitalization time.

## Conflicts of Interest

Dr. Bell receives compensation as an employee of illumiSonics Inc.. Dr. Bell, Dr. Dinakaran, Dr. Mackey, and Dr. Haji Reza have financial holdings in illumiSonics Inc..  All other authors have no conflicts of interest.

## Acknowledgments

The authors acknowledge Professor Paul Fieguth for his constructive discussions on image formation.  We thank Benjamin Ecclestone for his work on ethics requirements.  Finally, we thank Jean Flanagan at the University of Waterloo animal facility for her work in preparing the murine specimen.

# Reflection-mode virtual histology using photoacoustic remote sensing microscopy

# Supplementary Information

**Kevan Bell**[1,2,†], **Saad Abbasi**[1,†], **Deepak Dinakaran**[2,3], **Muba Taher**[4], **Gilbert Bigras**[5], **Frank K.H. van Landeghem**[6], **John R. Mackey**[3], **Parsin Haji Reza**[1*]

1. PhotoMedicine Labs, Department of Systems Design Engineering, University of Waterloo, Waterloo, Ontario, N2L 3G1, Canada
2. illumiSonics, Inc., Department of Systems Design Engineering, University of Waterloo, Waterloo, Ontario, N2L 3G1, Canada
3. Department of Oncology, University of Alberta, Edmonton, Alberta, T6G 1Z2, Canada
4. Division of Dermatology, Department of Medicine, University of Alberta, Edmonton, Alberta, T6G 2V1, Canada
5. Department of Laboratory Medicine and Pathology, University of Alberta, Edmonton, Alberta, T6G 2V1, Canada
6. Faculty of Medicine & Dentistry – Laboratory Medicine & Pathology Department, University of Alberta, Edmonton, Alberta, T6G 2B7, Canada

†. Equal contributions.

*Corresponding Author:* phajireza@uwaterloo.ca

## 1. Ultraviolet DNA/Nuclear Contrast

Previous investigations in photoacoustic imaging of cell nuclei have primarily used UV-C wavelengths typically around 266 nm [1] which aim to target the 260 nm absorption peak of DNA. However, 266 nm may not necessarily provide the greatest contrast against surrounding tissue regions. This disconnect between maximizing for absolute signal strength (i.e. targeting the absorption peak) and maximizing contrast may result from the non-zero UV contributions from neighboring chromophores such as other macromolecules in the nucleus and extranuclear structures of the cells, in the case of FFPE samples, the background paraffin. As such, we sought the optimal wavelength for nuclear contrast.

A 1 mg/mL solution of DNA (DNA from human placenta, Millipore Sigma) was prepared and placed directly onto the UV window of the tunable PARS. The excitation wavelength was then swept to produce a PARS-based optical absorption spectrum. This is shown in Figure SI1a. The overall shape is similar to that reported by more conventional spectroscopy techniques [2], with only a small shift in the locations of the peak and trough inside the UV range. It suggests that the 260 nm-range may be optimal for providing DNA signal strength, however, as will be shown in the next experiments this is not necessarily optimal for recovery of nuclear structure within tissues, which may be explained by the presence of other nuclear structures aside from DNA, such as histones, ribosomes and other specialized nuclear proteins [3].

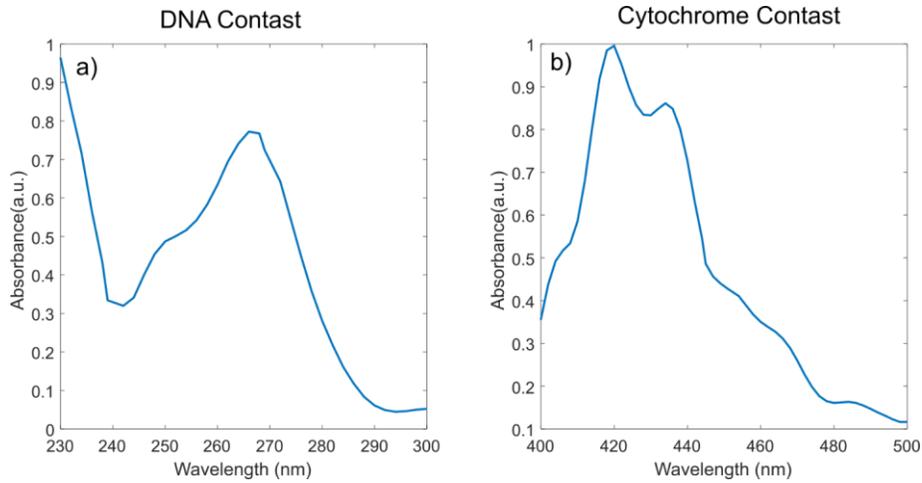

**Fig. SI1** Optical absorption spectra captured by the tunable PARS system for (a) DNA and (b) Cytochrome solutions.

A study was conducted to explore optimal excitation wavelengths for extracting the nuclear structure from FFPE samples. The study consisted of imaging the same region of a FFPE breast slide with multiple excitation wavelengths. An example of the results from this are shown in Figure SI2. Similar to previously reported studies on this topic looking at contact-based photoacoustic contrast, it was found that 250 nm appeared to give the best contrast for nuclei against the background. For this reason, the two-color PARS used 250 nm for DNA contrast.

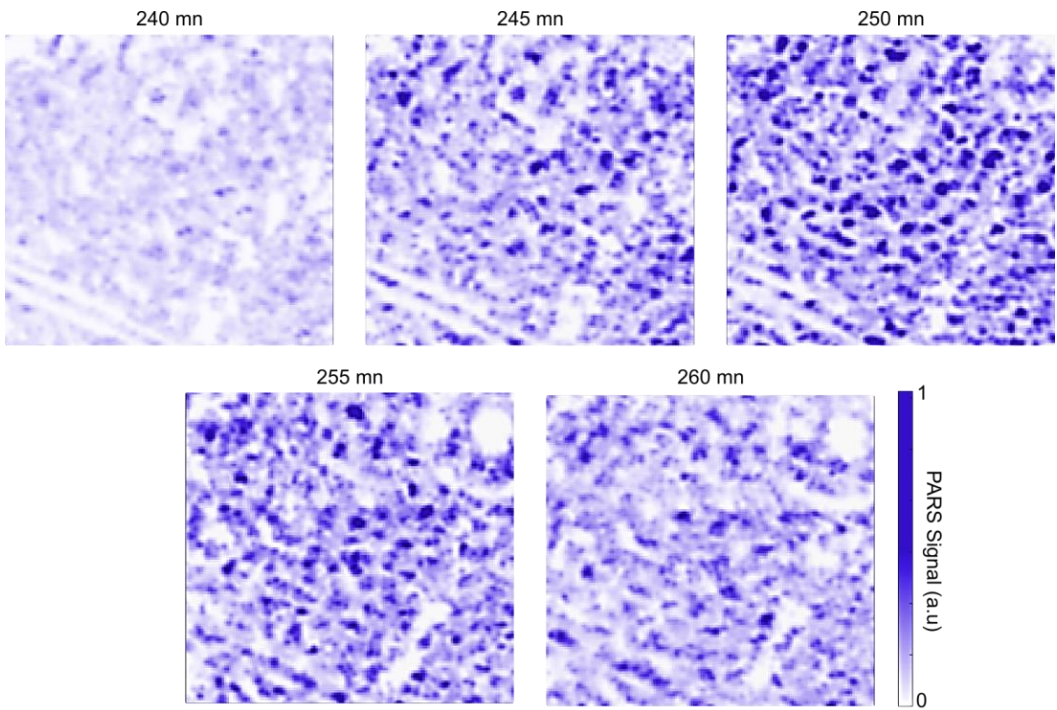

**Fig. SI2** A summary of a nuclear contrast study performed on the tunable PARS system.

## 2. Blue/Green Hemeprotein Contrast

Conventional H&E tissue preparations highlight both nuclear contrast along with that of the surroundings cytoplasm.  With the UV excitation providing nuclear structure, cytoplasm may be recovered by targeting hemeproteins such as cytochromes. These hemeproteins are known to provide similar absorption spectra to that of hemoglobin; however, they appear in significantly lower concentrations within the surrounding tissue as compared to hemoglobin within erythrocytes. Cytochromes offer a strong absorption peak around 420 nm with a smaller peak in the mid 500 nm range [4]. PARS has previously recovered erythrocytes with excitation operating at 532 nm [5], but this wavelength was inappropriate for cytoplasm recovery as it required high excitation pulse energies. As with the nuclear contrast, we sought the optimal excitation wavelength for cytoplasm recovery looking at reduced cytochrome C samples.

A 10 mg/mL solution of cytochrome (Cytochrome C from bovine heart, Millipore Sigma) was prepared and placed directly onto the UV window of the tunable PARS. The excitation wavelength was then swept to produce a PARS-based optical absorption spectrum.  This is shown in Figure SI1b. As with the DNA samples, the overall shape is again similar to that reported by more conventional spectroscopy techniques, showing a mid 400 nm absorption peak.  Since this value appears to be appropriate for extracting cytoplasm morphology, it was used directly. An excitation wavelength of 420 nm was selected for the two-color PARS to extract cytoplasm structure.

## 3. System Layout

The two PARS architectures used in this work are shown in detail in Figure SI1.  In both systems, care must be taken to minimize achromatic effects due to the wide range of wavelengths each system uses. The two-color tunable PARS can operate with an excitation ranging from 210 nm to 680 nm.  However, the poor beam quality which results from the optical parametric amplification inside the source is inappropriate for creating near-diffraction-limited foci and is highly astigmatic and elliptical. This requires spatial filtering of the excitation pulses before they can be fed into the imaging head. Since this task was implemented using refractive optics, the total wavelength range needed to be separated into smaller regions which would work with achromatic optical subsystems. Two of the arms, the UV arm (pink in Fig. SI1) which went from 210 nm to 400 nm and the blue arm (blue in Fig. SI1) which went from 400 nm to 505 nm used a pair of achromatic air-spaced doublets and a 50 µm pinhole. Meanwhile, the green-red arm (green in Fig. SI1) which ranged from 505 nm to 680 nm used a single-mode fiber. These spatial filters removed much of the undesired astigmatism from the beams which were then joined with the 1310 nm detection. The excitation beam quality produced by the picosecond 266 nm used in the single-color PARS was of sufficient quality to be used directly and thus did not require any spatial filtering.  However, some residual 532 nm needed to be removed which was accomplished by a prism.  In both systems, this wide range of wavelengths was focused on the sample using reflective objective lenses.  These are selected to avoid any additional chromatic effects and because such large wavelength ranges, ranging from UV to SWIR, are not typical of standard off-the-shelf refractive systems.

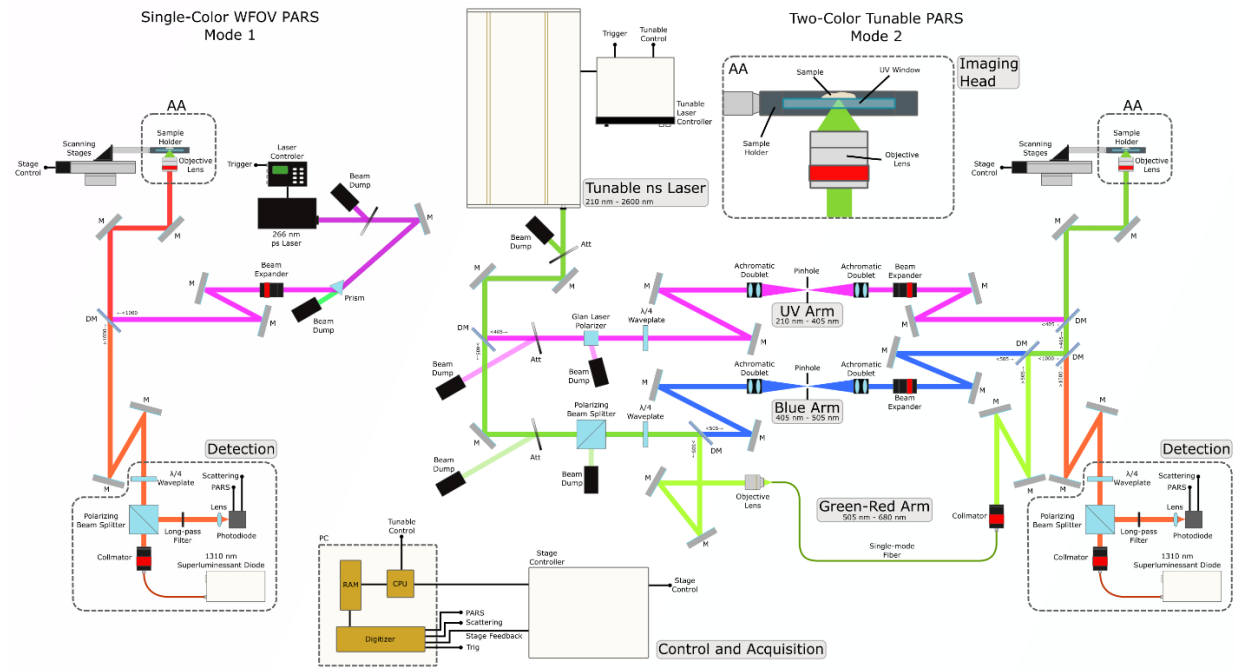

**Fig. SI3** An overview of the PARS systems used in the manuscript. The two-color tunable PARS is on the left. Excitation is provided by an Ekspla NT242 series nanosecond pulsed source. This beam is then split into three wavelength ranges (210 nm to 405 nm, 405 nm to 505 nm, and 505 nm to 680 nm) so that beam astigmatism and elipsisity may be improved through three spatial filters. These three paths are combined with the detection path, all of which are fed into the imaging head. The single-color WFOV PARS is shown on the right. It features a Bright Solutions picosecond pulsed source which emits primarily 266 nm but also contains some residual 532 nm. This 532 nm is removed using a dispersing prism. The excitation beam is then combined with the detection and fed into the imaging head. The control and acquisition architecture is highlighted in the bottom left. It features a digitizer which collects feedback information from the scanning stages, PARS signals, scattering signals, and trigger signals. This information is then streamed directly to the system RAM where it can be processed into images. Components are abbreviated as: Mirror (M), attenuator (Att.), and dichroic mirror (DM).

## 4. Imaging Speeds and field of views

The PARS systems used in this work present two unique imaging modes. The Single-color mode acquisitions are produced using a 266 nm source operating at 50 kHz pulse repetition rate and a set of 2D linear-drive scanning stages. This apparatus can acquire images which are several centimeters in size with down to around 1 µm lateral sampling. Here, large images can be formed which may be useful for rapid gross assessment of samples with the presented architecture capable of forming 13 mm x 13 mm at a 4 µm lateral sampling in roughly 15 minutes. The latter a resolution of this system is wholly limited by lateral sampling size. However, the particular stages used in this system did not provide sufficient velocity stability for steps smaller than around one micron. As will be further described of the resolution section, this provided a lateral the solution of 2.52 ± 0.44 µm suggesting that the system resolution is limited by the stages as the single-color WFOV utilizes a 0.3 NA objective lens which would otherwise provide higher lateral resolution.

The tunable two-color architecture used to acquire full false-color H&E-like visualizations uses a much slower pulse repetition rate laser which can only operate up to 1 kHz. For this work, the system has been optimized for performance at the two characteristic wavelengths, 250 nm and 420 nm. Images are acquired using 2D screw-driven scanning stages. Although these are capable of wide scans, a reasonable limit is imposed due to the relatively slow interrogation rate of the system. For this reason, individual tunable frames are limited to 1.6 mm by 1.6 mm with 900 nm which are acquired in 54 minutes per wavelength. As with the stage scanning on the WFOV modality, lateral resolution is primarily limited by lateral sampling of the stages.

## 5. Resolution Study

Resolutions studies were carried out on both the single-color and two-color PARS systems. These were conducted by imaging a 200 nm gold nanoparticles suspension. Since the spatial sampling of each system was assumed to be larger than the optical resolution, treating the visualized particles as ideal point-spread functions did not seem appropriate. In lieu of this, a center-to-center metric was used an average across multiple close particles for each wavelength of each system. The results of the study are summarized in Figure SI2. Resolution measurements indeed appeared to be limited by the accuracy of the scanning stages. Given the 0.5 NA objective used on the two-color system we would expect the lateral resolutions to be around 305 nm and 512 nm for 250 nm and 420 nm respectively using the $Res = 0.61\,\lambda/NA$ metric. However, the measured lateral resolution values were much closer to those expected by Nyquist spatial sampling, which would provide ~2 µm for 266 nm and ~600 nm for 250 nm and 420 nm. Note that this does not speak to the positional accuracy of the scanning stages, but rather their velocity accuracy.

The WFOV PARS used a linear-drive set of scanning stages which could not provide rapid positional feedback, and thus reconstruction relied solely on keyed positional information with an assumption of constant velocity between. However, this velocity varied by several percent between keyed locations resulting in positional accuracy for below that reported for the stages. More direct feedback methods must be investigated to improve this aspect. The two-color PARS used a screw-driven set of scanning stages, which although substantially slower, provided substantially improved velocity consistency, thereby improving positional accuracy upon reconstruction.

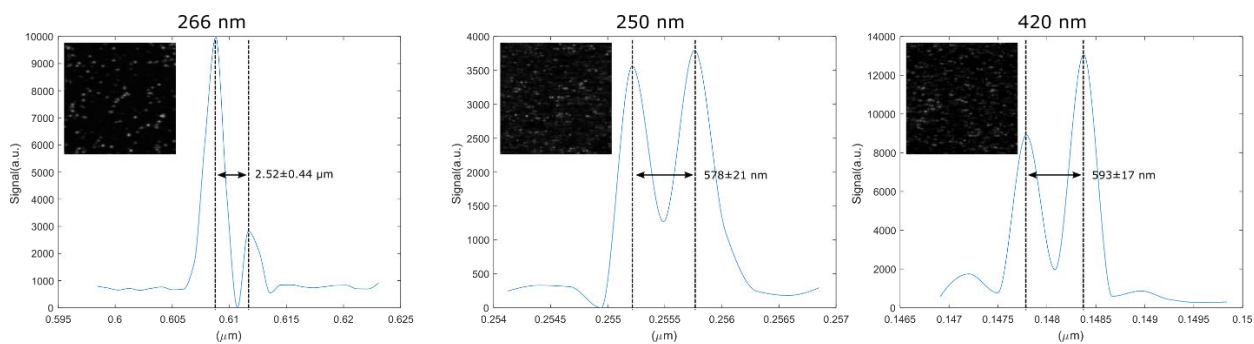

**Fig. SI4.** Resolution study for tunable architecture. Resolutions are taken as center-to-center distances between 200 nm gold nanoparticles. On the single-color WFOV system lateral resolution is found to be 2.52 ± 0.44 µm. On the two-color system values are found to be 578 ± 21 nm for 250 nm excitation and 593 ± 17 nm for 420 nm excitation.

## 6. SNR Study

A study of the single-color (mode 1) and the two-color (mode 2) signal-to-noise ratio (SNR) performance is presented. Signal to noise (SNR) ratios are calculated from various data sets using the following definitions for SNR$_{max}$ and SNR$_{mean}$

$$SNR_{max} = 20log\left(\frac{Max(S_i)}{\sigma_n}\right)$$

$$SNR_{mean} = 20log\left(\frac{\acute{S_i}}{\sigma_n}\right)$$

where $S_i$ is a collection of image pixels denoted as signal, and $\sigma_n$ is the standard deviation of a collection of image pixels denoted as noise. Values for the single-color and the two-color system for the four sample types are presented in Tables SI1 and SI2. As one example from this study, the max SNR on FFPE slides of human skin was measured to be roughly 55 dB using 20 nJ pulses in 266 nm on the single-color system. For another example, the max SNR on FFPE slides of human skin was measured to be roughly 41 dB using 35 nJ pulses in 250 nm and 26 dB using 120 nJ pulses in 420 nm on the two-color system.

**Table SI1**

| Single-Color PARS SNR | | | |
|---|---|---|---|
| **Sample Type** | **Wavelength (nm)** | **Max (dB)** | **Mean (dB)** |
| FFPE Slides | 266 | 55.0 ± 3.6 | 38.0 ± 5.6 |
| FFPE Blocks | 266 | 55.2 ± 0.9 | 34.7 ± 2.6 |
| Frozen Pathology | 266 | 53.7 ± 12.5 | 42.9 ± 7.1 |
| Fresh | 266 | 53.6 ± 1.5 | 38.4 ± 2.5 |

**Table SI2**

| Two-Color PARS SNR | | | |
|---|---|---|---|
| **Sample Type** | **Wavelength (nm)** | **Max (dB)** | **Mean (dB)** |
| FFPE Slides | 250 | 41.2 ± 1.6 | 25.9 ± 1.4 |
| | 420 | 34.7 ± 7.1 | 16.3 ± 1.7 |
| FFPE Blocks | 250 | 46.2 ± 4.3 | 24.5 ± 5.8 |
| | 420 | 31.1 ± 2.7 | 11.8 ± 1.4 |
| Frozen Pathology | 250 | 47.4 ± 5.4 | 36.4 ± 4.4 |
| | 420 | 51.6 ± 1.9 | 32.3 ± 22.7 |
| Fresh | 250 | 44.6 ± 5.9 | 24.5 ± 7.3 |
| | 420 | 46.5 ± 4.2 | 26.7 ± 9.0 |